\documentclass[12pt]{article}
\usepackage[american,british]{babel}
\usepackage{graphicx}
\usepackage[width=18cm,height=25cm,top=2cm,noheadfoot]{geometry}

\font\bb=cmbx10 scaled \magstep2
\begin{document}

\title{Quantum Monte Carlo Study of electrons in low dimensions}
\author{A. Malatesta and Gaetano Senatore\\
{\it INFM and Dipartimento Di Fisica Teorica dell'Universit\`a di Trieste}\\
{\it Strada Costiera 11, I-34014  Trieste} }
\date{}
\maketitle
\selectlanguage{british}
\thispagestyle{empty}

\begin{center}
\parbox{15cm}{\small {\bf Abstract.}  We report on a diffusion Monte Carlo
 investigation of model electron systems in low dimensions, which should
be relevant to the physics of systems obtainable nowadays in semiconductor
heterostructures. In particular, we present results for a one dimensional
electron gas, at selected values of the coupling strength and confinement
parameter, briefly analyzing the pair correlations and relating them to 
predictions by Schulz for a Luttinger liquid with long--range interactions.
We find no evidence of the the Bloch instability yielded by approximate
treatments such as the STLS and DFT schemes. }
\end{center}

\bigskip

\section*{\bb 1. INTRODUCTION}

Electrons in low dimensions, as found in modern semiconductor
devices\cite{nano89}, are greatly affected by correlations effects that may
dramatically change their behavior and bring about new phenomena and phase
transitions. However, at zero magnetic field and in simple systems, such as
layers or wires, very low densities are necessary for correlations to play an
important role. The situation is somewhat better in coupled systems. The
additional correlations due to the interlayer (or interwire) interactions may
help in pushing transitions to larger and more easily accessible
densities\cite{fs97}, yielding richer phase diagrams and possibly new
phenomena, such as superconductivity or excitonic
condensation\cite{degs99}. In all cases, the quantum Monte Carlo technique
provides an effective tool, which allows the determination of the static
properties of these systems with unprecedented accuracy\cite{dmcegas,sm94}.
While in the invited paper delivered at the St-Malo by one of us (GS) results
were presented for both an electron-hole bilayer and a model quantum wire,
due to lack of space here we shall restrict to the latter system. A detailed
account of the electron--hole bilayer simulations will be given
elsewhere\cite{degs00}.

Perhaps, the theoretical interest for one-dimensional (1D) models is due in
part to their inherent simplicity, which often results in exact
solutions\cite{mattis93}. In fact, the problem of interacting Fermions
simplifies in one dimension and one can show that the familiar concept of
Fermi liquids has to be abandoned in favor of that of Tomonaga-Luttinger
liquids\cite{lutt}. The interest in 1D models has grown even bigger in recent
years, thanks to the advances in fabrication techniques and the realization of
the so called quantum wires\cite{nano89,yacob,thomas9699}, i.e.,
quasi-one-dimensional electron systems. Thus, the investigation of model 1D
electron gases with numerical simulations, which yield results of high
accuracy if not exact, is particularly appealing---both in relation to
experiments and to other theoretical approaches.  

\section*{\bb 2. THE MODEL}
    
In a quantum wire the electronic motion is confined in two directions (say y
and z) and free in the third one, x. In the simplest approximation, one
assumes that the energy spacing of the one-particle orbitals for the
transverse motion is sufficiently large, so that only the orbital lowest in
energy, say $\phi(y,z)$, needs to be considered. Hence the total wavefunction
of the many-electron system will factorize in a irreducible many body term for
the x motion, $\Psi(x_1,\, ...,\, x_n)$, times a product of $\phi$'s, one per
particle.  Tracing out the transverse (y,z) motion from the full Schr\"odinger
equation yields an effective 1D problem with an effective 1D interparticle
potential. Evidently, different models of confinement yield different
effective potentials. One of the firt models assumes a harmonic confinement
in the transverse plane\cite{model}. More recently a hard wall confinement has
been investigated, with the electrons moving freely in a cylinder of given
radius\cite{modelf3}. One may also start from the 2D electron gas and apply a
confining potential in one direction; again, both harmonic\cite{modelh2} and
hard wall\cite{modelf2} confinements have been considered.

Here, we choose the model of Ref. \cite{model}, with a harmonic confining
potential $ U_c({\bf r})=(\hbar^2 /8m^{*}b^4)(y^2+z^2)$ and a coulombic
electron-electron interaction $e^2/\epsilon r$. The resulting effective 1D potential is readily shown to be
$v(x)=(e^2/\epsilon)(\sqrt{\pi}/2b)\exp [ (x/2b)^2]{\rm erfc}[|x|/ 2b]$,
with  Fourier transform
$v(q) = (e^2/\epsilon)E_1[ (bq)^2] \exp[ (bq)^2].$
Above $m^*$ and $\epsilon$ are, respectively, the effective mass of the
carriers and the dielectric constant of the medium in which the carriers move,
and $b$ measures the wire {\it width}.  One can easily check that $v(x)$ is
long ranged, with a Coulomb tail $ e^2/\epsilon |x|$, and is finite at the
origin, $v(0)=(e^2/\epsilon b)(\sqrt{\pi }/2)$. The 1D system is made neutral
by introducing a background that cancels the $q=0$ component of the pair
interaction.\cite{model}.

Earlier investigations of this model have employed the so-called STLS
approximation\cite{stls}, either in its original version\cite{model} or in its
sum rule approach\cite{camels97}. Both the paramagnetic\cite{model,camels97}
and the ferromagnetic\cite{ferro} phases have been studied and the occurrence
of a Bloch instability [transition from the paramagnetic to ferromagnetic
state] has been predicted\cite{ferro}. This, according to the authors, could
explain the {\it anomalous} plateau which has been observed in the conductance
of  GaAs quantum wires, in the limit of single channel
occupancy\cite{thomas9699}.

\section*{\bb 3. DMC RESULTS}
                                                   
To study our model quantum wire we resort here to fixed-node diffusion Monte
Carlo (DMC) simulations\cite{dmcf}.  As the exact nodes are known in 1D, for a
given number of particles we obtain exact estimates of the energy, within the
statistical error---the small systematic time step error involved with the
imaginary time integration was in fact extrapolated out\cite{amat}. The
estimates of other properties, such as static correlation functions and
momentum distributions, remain approximate though very accurate. Below, we
present some of our results for the energy and the structure, skipping
completely the technical details of our calculations which can be found
elsewhere\cite{amat,ama}.

We have performed simulations for three different values of the wire width,
$b/a_B^*=4,\, 1,\, 0.1$, at selected values of the coupling parameter $r_s$,
defined in 1D by $\rho=N/L \equiv1/2 r_s a_B^*$. Here,
$a_B^*=\hbar^2\epsilon/m^*e^2$ is the effective Bohr radius of the material.  We
should remark that for the model at hand the coupling strength, defined as the
ratio between the potential energy of a pair of particles at the mean distance
$r_sa_B^*$ and the Fermi energy, is proportional to $r_s \times[r_s a_B^*
v(r_sa_B^*)]=r_s \times f(r_sa_B^*/b) $, with $f(x)$ a growing function of $x$.
Thus, at fixed $b$, the coupling actually increases more than linearly with
$r_s$, whereas at given $r_s$ the coupling increases with decreasing
$b$---reflecting the obvious fact that a narrower wire enhances the effect of
the Coulomb repulsion.

\subsection*{\bb 3.1 The Energy}
 
\begin{figure}
\begin{center}\leavevmode
\hspace{-0.3cm}
\includegraphics[scale=.47]{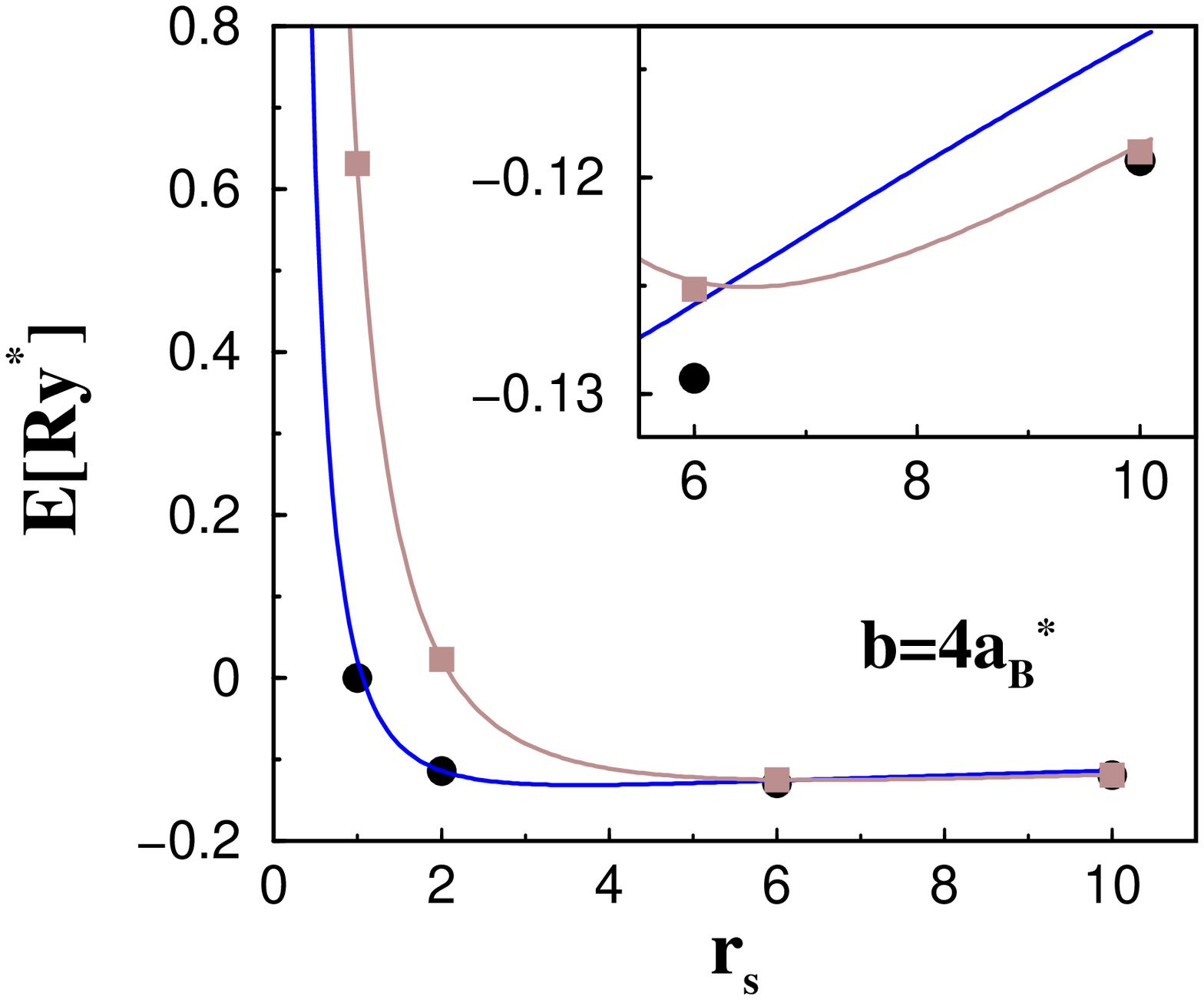}
\hspace{0.7cm}
\includegraphics[scale=.47]{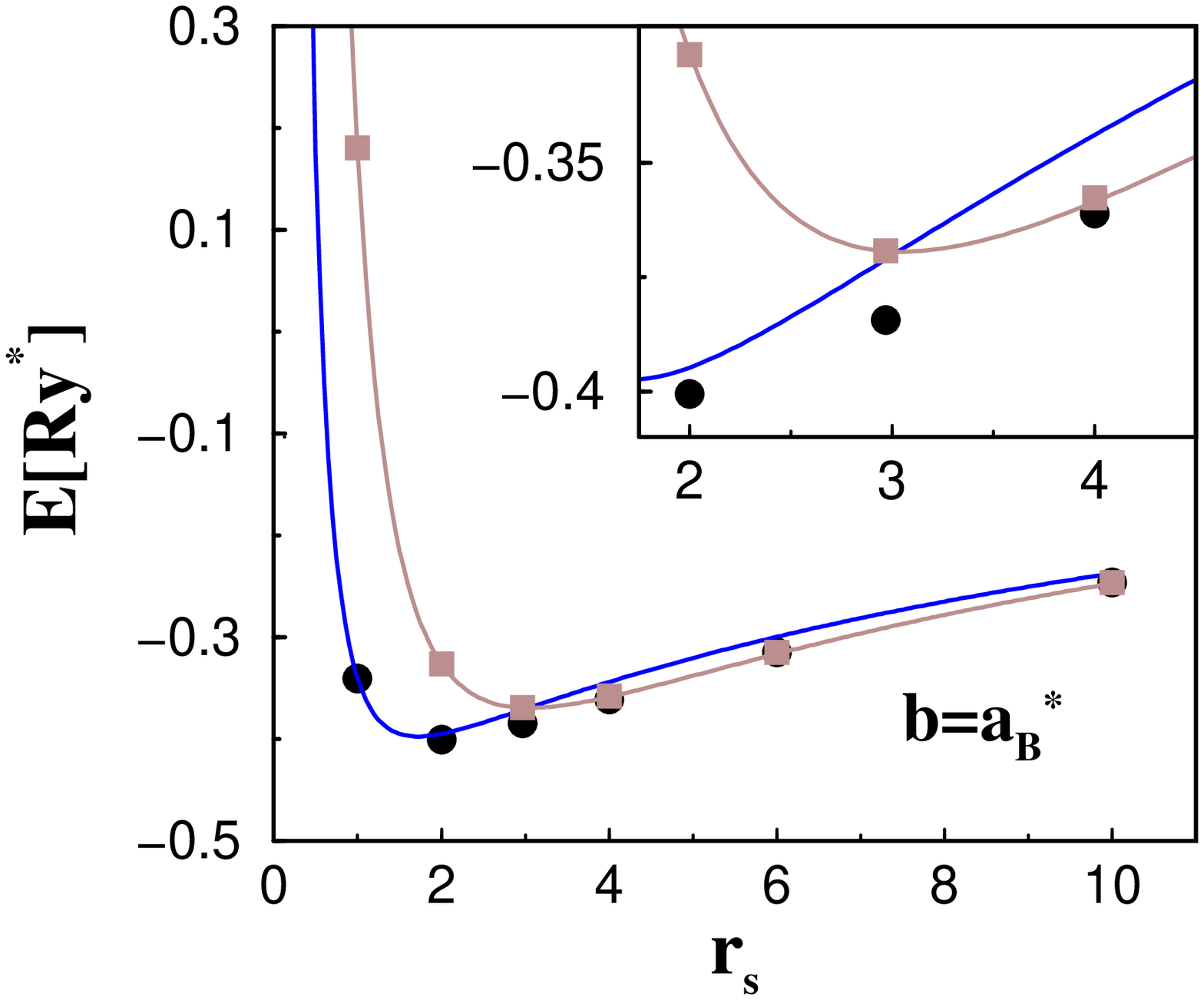}
\caption{\small DMC ground state energy per particle, in
$Ry^*=e^2/2\epsilon a_B^*$, of the paramagnetic (squares) and ferromagnetic
(circles) fluids, in the thermodynamic limit $N=\infty$. The error bars are
much smaller than the symbols. The predictions of STLS are given by the black
and gray line, respectively for the paramagnetic and ferromagnetic phase. The
insets show the Bloch instability yielded by the STLS scheme but not by our
DMC simulations.}\label{en}
\end{center}\end{figure}

Our DMC ground state energies for wires with $b=4a_B^*$ and $b=a_B^*$ are
shown in Fig. \ref{en}, together with the results of the STLS scheme, which is
easily solved numerically. We should note here that the alternative sum rule
approach to STLS yields results\cite{camels97} somewhat different from the
present ones, in terms of correlation energy\cite{foot}. However, as the
correlation energy is a small fraction of the ground state energy, such
differences can be neglected here and the curves in Fig. \ref{en} can also be
taken as representative of the results of Ref. \cite{camels97}.
It is evident, from Fig. \ref{en}, that (i) the STLS predicts a transition
from the paramagnetic to the ferromagnetic phase as the coupling strength is
increased and (ii) the {\it critical} $r_s$ value of such a transition
increases with $r_s$\cite{ferro}. Our exact energies, on the contrary, give
the correct ordering of the two phases imposed by the Lieb--Mattis
theorem\cite{lm}: the ferromagnetic fluid is higher in energy than the
paramagnetic one. In fact, the distance in energy of the two phases closes up
with increasing $r_s$ and, at the largest values of $r_s$ considered here,
falls within the combined error bars of the two phases.  This is hardly
surprising. At large coupling, the strong correlations that build up in the
system keep particles well apart and thus the statistics ends up having
negligible effects on the total energy. In turn, this makes the sampling of
spin correlation extremely hard\cite{amat,ama}.

An additional comment, which is naturally prompted by Fig. \ref{en}, is that
at intermediate and large coupling the STLS performs much better for the
ferromagnetic phase than for the paramagnetic one. This is most easily
appreciated by looking at the insets in the figure. In fact, one may argue on
general grounds that it is easier to describe the fully spin polarized phase,
as part of the correlation is automatically built in by the symmetry
constraints.
      
\subsection*{\bb 3.2 The Structure}

\begin{figure}
\begin{center}\leavevmode
\hspace{-0.3cm}
\includegraphics[scale=.4]{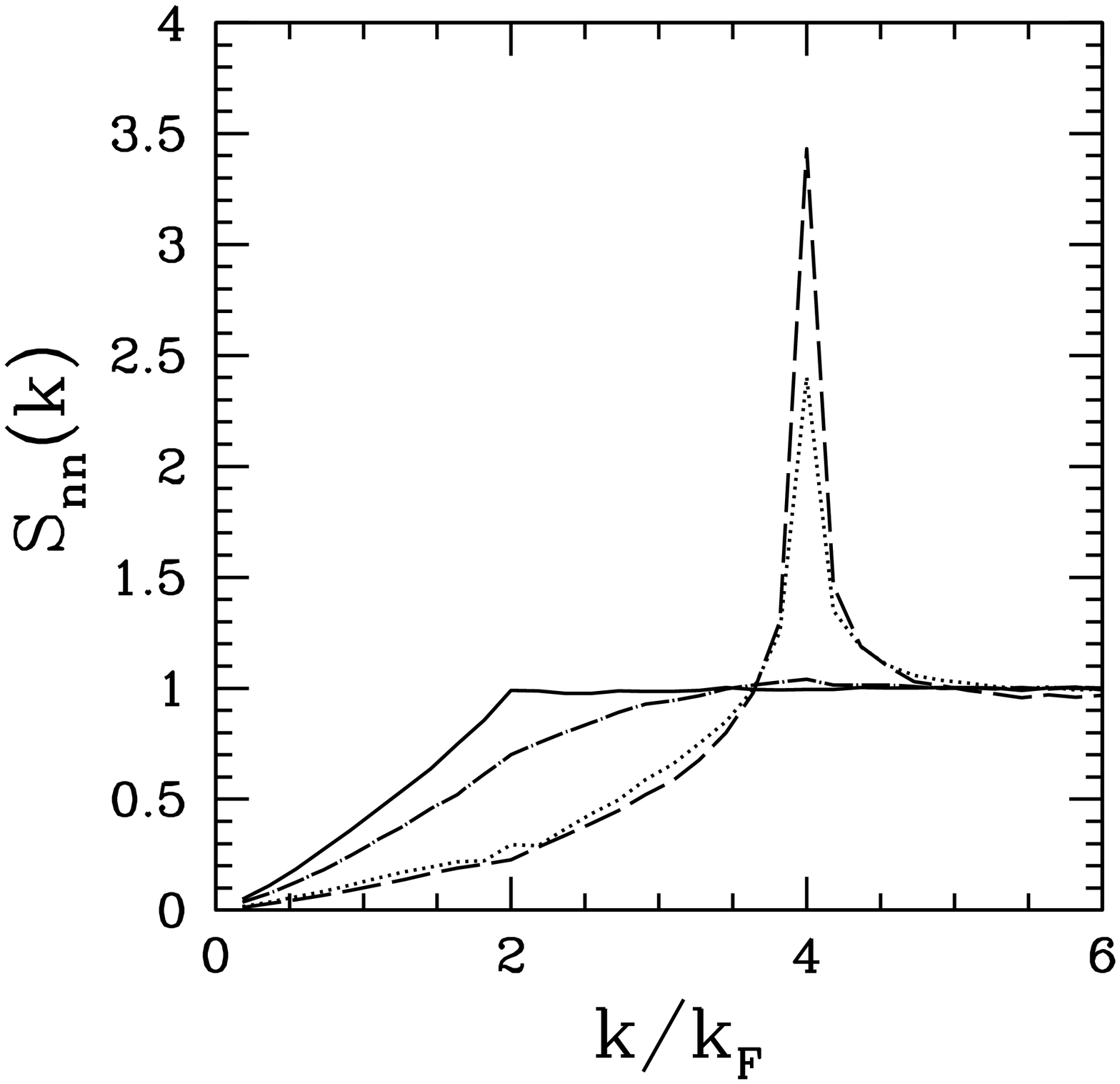}
\hspace{0.5cm}
\includegraphics[scale=.435,trim=0 -10 0 0]{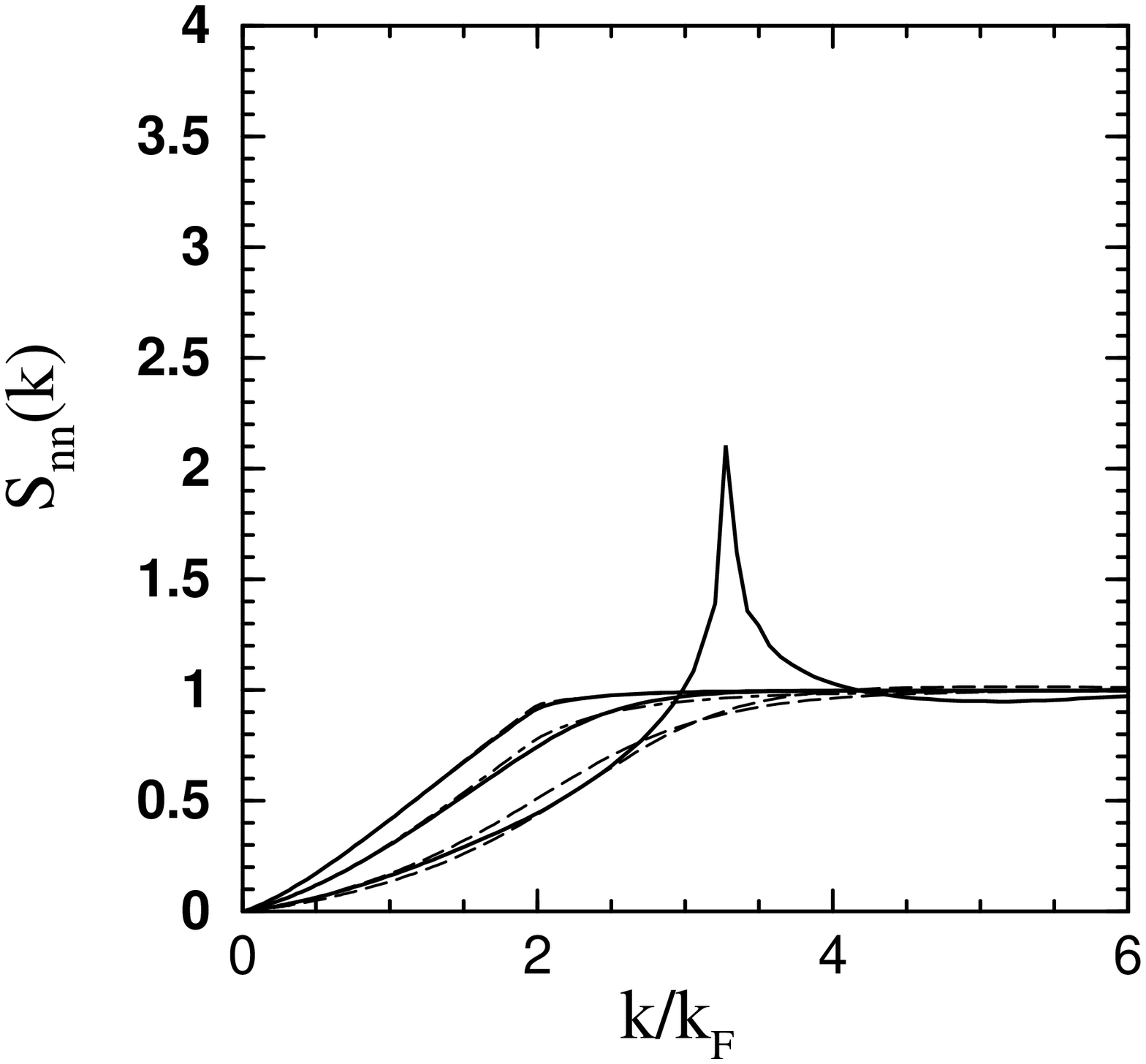}
\caption{\small Static structure factor of the paramagnetic fluid. The left
panel gives extrapolated DMC estimates for 22 particles and $b=a_B^*$, at
$r_s=1,\,2,\,6,\,10$; the errors are not visible on this scale. The right
panel gives the predictions of STLS ( dashed curves) and DSTLS (full
curves). In all cases a decreasing slope at the origin corresponds to
increasing $r_s$. Also, for the DSTLS only results for $r_s$ up to 6 are
shown.}\label{sk}
\end{center}\end{figure}
In a quantum wire interactions are enhanced, due to the reduced
dimensionality, and strong ordering may thus arise at large coupling, even
though genuine crystalline order is generally forbidden in 1D. In fact a 1D
system with an interparticle potential decaying as $1/|x|$ is borderline, in
this respect\cite{schulz93}.  Ordering may be characterized in terms of
structure factors, which measure ground state correlations between Fourier
components of one body densities. Thus, we shall focus on the number and
magnetization static structure factors, respectively $S_{nn}(k)$ and
$S_{mm}(k)$, which are defined by $ S_{\alpha \beta}(k)=\langle
\rho_{\alpha}(k) \rho_{\beta}(-k)\rangle/N$, with the number density
$\rho_n(k)=\rho_{\uparrow}(k)+\rho_{\downarrow}(k)$ and the magnetization
density $\rho_m(k)=\rho_{\uparrow}(k)-\rho_{\downarrow}(k)$. The (cross)
charge-spin correlations, measured by $S_{nm}(k)$, need not be considered
since they exactly vanish in the paramagnetic fluid.

The building up of a quasi-crystalline order with increasing the coupling is
clearly seen in our DMC results shown in Fig. \ref{sk} for $b=a_B^*$. The
static structure factor $S_{nn}(k)$, while very close to the Hartree-Fock
prediction at $r_s=1$, with increasing $r_s$ develops a pronounced peak at
$4k_F$, which in fact may be shown to be divergent with the number of
particles $N$, for large couplings, (see below). A pronounced peak at $4k_F$
corresponds in real space to slowly decaying oscillations, with period equal
to the average interparticle distance $2r_sa_B^*$, thus suggesting
quasi-crystalline order.  In the same figure we also give the predictions of
approximate theories such as STLS or its dynamical version\cite{dstls}
(DSTLS). The STLS only gives the lowering of $S_{nn}(k)$ at small and
intermediate values of $k$, for increasing $r_s$, but fails completely in
yielding a peak.  On the contrary, the DSTLS prediction develops a peak, with
increasing the coupling, though its position is off by about 20\%; the height
of the peak happens to almost coincide with that of the DMC result at
N=22. Similar DSTLS results were recently obtained for a slightly different
model of wire\cite{bul98}. We should mention that at the time of writing we
were not able to obtain a solution to DSTLS for $r_s> 6$.
\begin{figure}
\begin{center}\leavevmode
\hspace{-0.3cm}
\includegraphics[scale=.43]{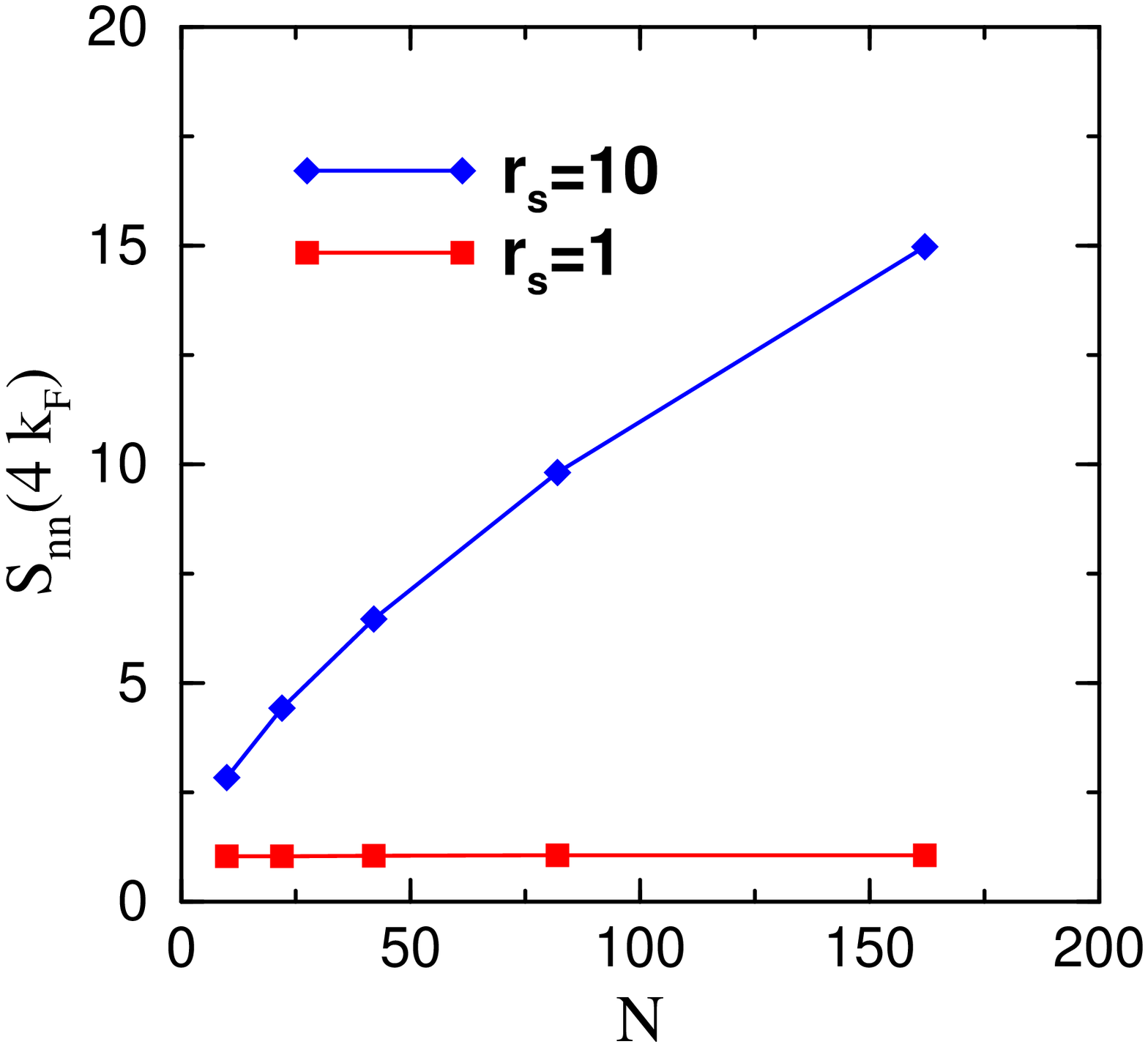}
\hspace{0.5cm}
\includegraphics[scale=.43]{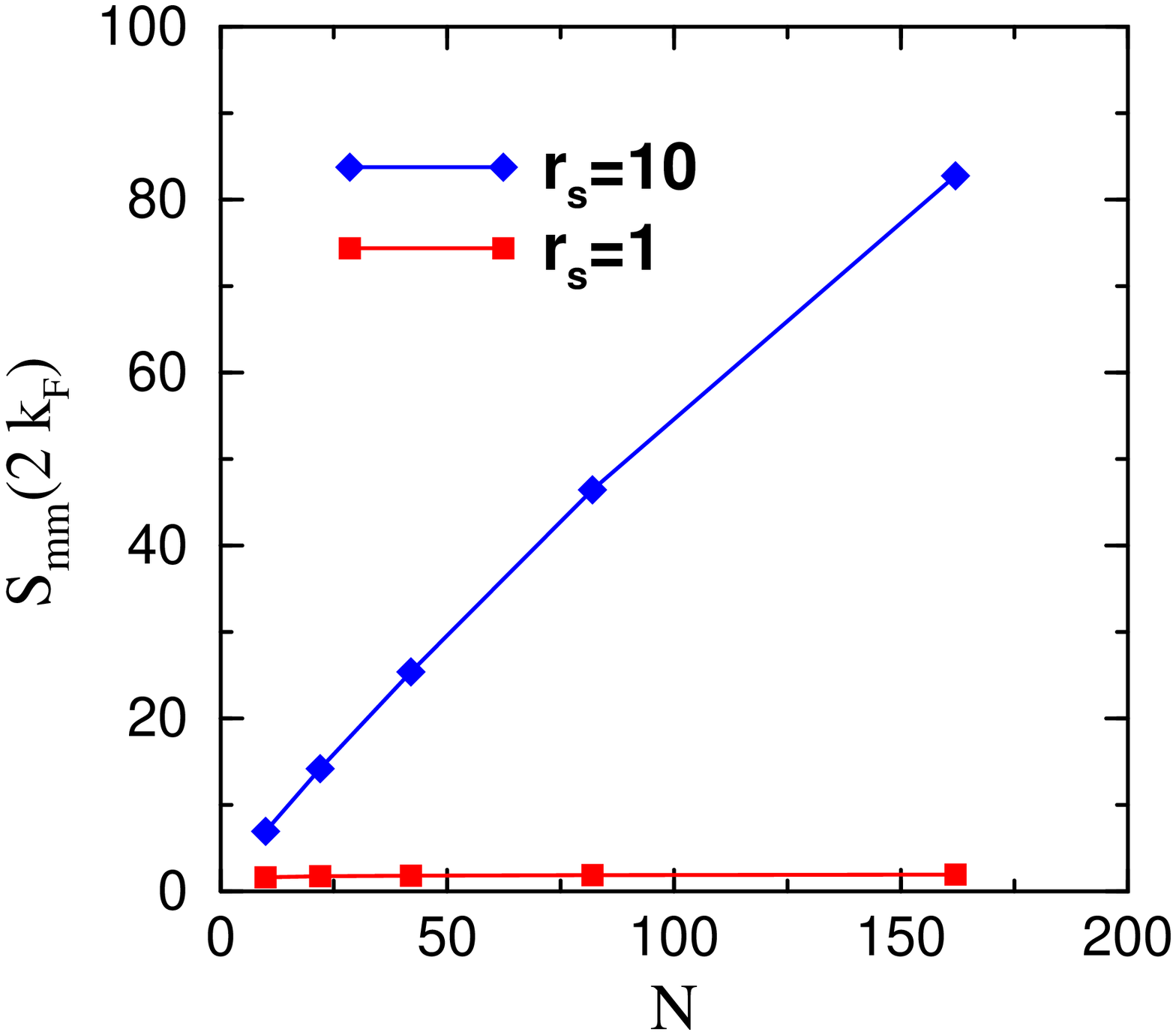}
\caption{\small Height of the peak at $4k_F$ in $S_{nn}(k)$ (left panel) and
of the peak at $2k_F$ in $S_{mm}(k)$ (right panel), for a wire with $b=a_B^*$,
as a function of the particle number $N$ and according to VMC. The squares and
diamonds correspond to weak and strong coupling, respectively.  }\label{peaks}
\end{center}\end{figure}

Recently, Schulz analyzed the properties of a yet different wire with long
range interactions also behaving as $e^2/\epsilon |x|$ at large $x$, resorting
to a linearized dispersion of the kinetic energy and employing the
bosonization technique\cite{schulz93}, which gives exact results for a a
Luttinger liquid\cite{lutt}.  He found persistent tails in the pair
correlations, both for the number and magnetization variables, implying a
divergent peak in the number structure factor $S_{nn}(k)$ at $4k_F$, and a
pronounced but finite peak in the magnetization structure factor $S_{mm}(k)$
at $2k_F$. In his prediction, however, real-space tails contain undetermined
interaction-dependent prefactors.

As we found that at $N=22$ our DMC and variational Monte Carlo (VMC) results
for the structure compare fairly well with each other\cite{amat,ama}, we have
employed VMC to study the $N$ dependence of the peaks of the structure
factors, which we shown in Fig. \ref{peaks}. In passing, we mention that our
VMC results for the paramagnetic fluid almost coincide with those obtained
from a harmonic treatment of a finite linear chain.  It is evident that, at
variance with the results for the Luttinger liquid, we have indication of
peaks diverging with $N$ only at large values of the coupling, but for both
$S_{nn}(k)$ and $S_{mm}(k)$. In addition $S_{mm}(2k_F)$ appears to grow faster
with $N$ than $S_{nn}(4k_F)$, again in contradiction with the results of
Ref. \cite{schulz93}. Possible explanation of these differences might be
traced to either the undetermined interaction dependent prefactors mentioned
above or to the fact that in the present study the full dispersion of the
kinetic energy was retained.

\subsection*{\bb 4. CONCLUSIONS AND ACKNOWLEDGEMENTS}

We have presented accurate results for one-dimensional electron gases adapted
to describe quantum wires of different width, focusing on the energy and the
pair correlations.  Our results for the energy, which are exact, do no involve
surprises: they satisfy the Lieb-Mattis theorem, in contrast to approximate
treatments\cite{camels97,bergreen96}, and rule out the occurrence of a Bloch
instability. Thus, the origin of the anomalous plateau observed in the
conductance of GaAs quantum wires, in the limit of single channel
occupancy\cite{thomas9699}, should be sought elsewhere.

Our results for the pair correlations, on the other hand, are intriguing. They
are not exact. Yet they should be rather accurate and is natural to make
comparison with the predictions for the Luttinger liquid studied by
Schulz\cite{schulz93}, which has a slightly different interparticle
interaction but with same long range tail. However, as we have observed above,
it does not appear possible to reconcile in a simple manner the predictions of
the present investigation with those of Ref. \cite{schulz93}. One possibility
could be that the unknown interaction-dependent constants entering the tails
of the pair correlations of the Luttinger liquid could in fact have a singular
dependence on the coupling.  At this time, we can only say that this issue
deserves further investigations, both with bosonization techniques, to fully
determine the coupling dependence of the tails in the pair correlations, and
with numerical simulations, to estimate structure factors in an exact
fashion. To this end one might resort to the recently proposed reptation Monte
Carlo\cite{rept}, which provides a simple direct way to evaluate ground state
averages of local operators exactly.

One of us (GS) is happy to acknowledge useful discussions with Saverio Moroni
and Allan H.  MacDonald. We should also thank  Stefania De Palo for reading the manuscript.


\begin{thebibliography}{99}
\bibitem{foot}For instance, for $b=a_B^*$ the discrepancies on the
correlation energy vary from 6\% (15\%) at $r_s=1$ to 1\% (11\%) at $r_s=10$
for the paramagnetic (ferromagnetic) phase, with the full numerical solution
of the STLS scheme yielding always more correlation energy. Yet, our STLS
results still underestimate the exact correlation energy for the paramagnetic
fluid (by 2\% at $r_s=1$ to 9\% at $r_s=10$) and for the ferromagnetic fluid
at small copling ( by 2\% at $r_s=1$), while overestimating the correlation
energy of the ferromagnetic fluid at larger values of the coupling (by a few
percent  for $4\leq r_s\leq 10$).
\vspace{-3mm}
\end{thebibliography}
\end{document}